# Nanometer-resolution 3D Tomographic and Vectorial Near-field Imaging in Dielectric Optical Resonators

Bingbing Zhu[1,2,†], Qingnan Cai[1,2,†], Yaxin Liu[1,2,†], Sheng Zhang[1,2,†], Weifeng Liu[1,2], Qiong He[1,2], Lei Zhou[1,2,*], Zhensheng Tao[1,2,*]

[1] State Key Laboratory of Surface Physics, Key Laboratory of Micro and Nano Photonic Structures (MOE), and Department of Physics, Fudan University, Shanghai 200433, China.

[2] Shanghai Key Laboratory of Metasurfaces for Light Manipulation, Fudan University, Shanghai 200433, China.

† These authors contributed equally to this work.

*Corresponding author. Email: phzhou@fudan.edu.cn; zhenshengtao@fudan.edu.cn.

## Abstract

All-dielectric optical nano-resonators, exhibiting exotic near-field distributions upon excitations, have emerged as low-loss, versatile and highly adaptable components in nanophotonic structures for manipulating electromagnetic waves and enhancing light-matter interactions. However, achieving experimental full three-dimensional characterization of near-fields within dielectric nano-resonators poses significant challenges. Here, we develop a novel technique using high-order sideband generation to image near-field wave patterns inside dielectric optical nano-resonators. By exploiting the phase-sensitivity of various harmonic orders that enables the detection of near-field distributions at distinct depths, we achieve three-dimensional tomographic and near-field imaging with nanometer resolution inside a micrometer-thick silicon anapole resonator. Furthermore, our method offers high-contrast polarization sensitivity and phase-resolving capability, providing comprehensive vectorial near-field information. Our approach can potentially be applied to diverse dielectric metamaterials, and becomes a valuable tool for comprehensive characterization of near-field wave phenomena within dielectric materials.

## Introduction

All-dielectric optical resonators have emerged as essential building blocks of nanophotonic structures for manipulating electromagnetic waves and enhancing light-matter interactions[1–4]. Different from their metallic counterparts, these resonators, composed of high-index dielectric and semiconductor materials, offer notable advantages, including very low absorptive losses, diverse resonances, and tunable responses through electrical doping or carrier-density gating[5]. Whether functioning as individual entities or within intricately patterned arrangements like metasurfaces[6,7], dielectric optical resonators have demonstrated a variety of high-efficiency and exotic electromagnetic responses, including polarization controls[8–11], ultrathin lenses[12–15], large photonic spin Hall effect[16], and nanoscale light confinement in air[17].

In stark contrast to plasmonic structures, where near-field effects are usually confined to the surfaces, the enhancement and manipulation of near-fields with dielectric resonators rely on deep penetration of incoming light into the resonator structures, inducing electric and magnetic Mie and/or Fabry-Pérot-like resonances[3,18–20]. Therefore, comprehensive experimental characterization of the three-dimensional (3D) distribution of near-field wave patterns is essential for evaluating design performances and verifying novel nanophotonic concepts. However, achieving 3D tomographic and vectorial imaging of optical near-fields inside thick dielectric materials poses significant challenges for concurrent near-field imaging techniques. Although scanning near-field optical microscopy, the most prevalent technique for optically-based near-field imaging, can achieve sub-10 nm spatial resolution in the transverse directions[21–27], its depth of detection is limited to ~100 nm due to the rapid decay of the evanescent field away from the surface[28,29](Fig. 1a). While alternative electron-based techniques hold great promise for high-spatial-resolution near-field imaging[30–36], their probing depth is typically less than 50 nm (Fig. 1b), making them incapable of probing fields deep inside dielectric materials.

In this context, nonlinear optical microscopy emerges as a promising approach, because nonlinear optical wave mixing can occur at significant depths within a dielectric material. While previous studies have harnessed nonlinear optical effects to enhance the contrast or resolution of near-field scattering[37–39], the mapping results are usually two-dimensional, acquired by integrating along the sample thickness. This limitation is primarily attributed to the narrow spectral bandwidth associated with commonly utilized nonlinear processes, such as second-harmonic generation[40–42], third-harmonic generation[43,44], or four-wave mixing[45,46], etc. High-order nonlinear optical processes, which can provide much richer information about spatial and temporal distribution of near-field wave patterns, have not been fully exploited yet.

In this work, by employing the high-order sideband generation process[47–49] for near-field optical microscopy, we demonstrate the first nanometer-resolution, 3D tomographic and vectorial imaging of optical near-field wave patterns within micrometer-thick dielectric resonators. The specific near-field modes of interest here are the mid-infrared anapole modes excited within silicon resonators, which represent an important class of dark-mode near-fields[50–52] that have been previously inaccessible beyond the resonator surfaces. We retrieve the key information pertaining to the longitudinal dimension by simultaneously imaging the near-fields with up to 6 harmonic orders covering a broad spectral range. The phase-sensitivity inherent in this process makes each harmonic order sensitive to near-fields at distinct depths, achieving a conditional super-resolution of ~130 nm in the longitudinal direction. In the transverse directions, a high spatial resolution of 920 nm is attained through the nonlinear optical effects. Furthermore, we demonstrate high-contrast polarization sensitivity and phase-resolving capability of our method, the latter of which is realized through quantum-path interference between neighboring harmonic orders[53].

As inherent optical nonlinearity is abundant in dielectrics or semiconductors, our approach can potentially be applied to a wide range of dielectric metamaterials. The powerful capabilities of high-order nonlinear optical microscopy position it as a valuable tool for characterizing nanophotonic devices and optimizing near-field patterns under diverse schemes of light-matter interactions.

## Results

**High-order sideband microscopy.** In general, high-order nonlinear optical processes utilized in nonlinear optical near-field microscopy arise from wave-mixing between a pump laser, which resonantly excites the near-fields within a dielectric resonator at a wavelength of $\lambda_{\mathrm{pu}}$, and a probe laser with a shorter wavelength $\lambda_{\mathrm{pr}}$ that is far off-resonance. The corresponding pump and probe laser frequencies are $\omega_{\mathrm{pu}}$ and $\omega_{\mathrm{pr}}$, respectively. The spatial and temporal near-field information deeply embedded within a dielectric optical resonator is encoded in the output beam at a wavelength of $\lambda_{(m,n)} = \frac{\lambda_{\mathrm{pu}}\lambda_{\mathrm{pr}}}{m\lambda_{\mathrm{pu}}+n\lambda_{\mathrm{pr}}}$, where $m$ and $n$ represents the number of photons from the probe and pump lasers, respectively, involved in the nonlinear process. This output beam is subsequently collected and analyzed using conventional far-field optics (Fig. 1c).

The transverse spatial resolution of the microscope is determined by $\lambda_{\mathrm{pr}}$. To resolve the longitudinal field distribution, we utilize light at different wavelengths with varying coherence lengths during the nonlinear wave-mixing process to extract near-field information at different depths. Specifically, the longitudinal resolution $\delta l_z$ is governed by the variation of the longitudinal wavevector mismatch $\delta k_z$ in the nonlinear optical process: $\delta l_z \propto \frac{2\pi}{\delta k_z}$. The value of $\delta k_z$ depends on the bandwidths of the output beam and the probe laser $\delta\lambda_{\mathrm{pr}}$, as well as the refractive indices of the dielectric medium. It can be approximately given by:

$$\delta k_z \approx \frac{A_{(m,n)}}{\lambda_{(m,n)}^2}\delta\lambda_{(m,n)} + m\frac{B_{(m,n)}}{\lambda_{\mathrm{pr}}^2}\delta\lambda_{\mathrm{pr}}, \tag{1}$$

where $A_{(m,n)}$ and $B_{(m,n)}$ are the coefficients associated with the refractive indices at $\lambda_{pr}$, $\lambda_{pu}$ and $\lambda_{(m,n)}$. The term $\delta\lambda_{(m,n)}$ represents the output bandwidth covered by varying the harmonic orders *n* (and *m*). A detailed derivation of Eq. (1) is provided in Methods. Consequently, high longitudinal spatial resolution can be achieved by employing a broad spectral bandwidth either in the output beam by varying *n* or in the probe laser.

In this study, we demonstrate this approach by investigating mid-infrared (MIR) anapole modes excited within silicon resonators at a resonant wavelength of $\lambda_{Pu}\approx 13.4$ μm (see SM Section S3). The probe laser operates at a wavelength of $\lambda_{Pr}\approx 967$ nm. For a single lowest-order nonlinear optical process (e.g. *m*=1 and *n*=2), the $B_{(m,n)}$ coefficient is very small, which would require an extremely broad $\lambda_{pr}$ spectrum for high-resolution tomographic imaging (see the discussions in Methods). To overcome this limitation, we employ multiple high-order nonlinear processes by varying *n* and *m*, also dubbed as high-order sideband generation (HSG).

In Fig. 1d, we present a typical HSG spectrum obtained from a silicon-cylindrical resonator, spanning more than an octave and centered around 550 nm. The HSG spectrum is characterized by the combination of *m* near-infrared (NIR) probe photons and *n* MIR near-field photons: $\omega_{(m,n)} = m\omega_{\mathrm{Pr}} + n\omega_{\mathrm{Pu}}$, where the near-field photons possess the pump-laser frequency $\omega_{\mathrm{Pu}}$. Here, *m*+*n* is an odd integer for materials with inversion symmetry[54]. In the following, we denote different HSG orders by (*m*, *n*). The spectrum in Fig. 1d consists of two regions with (*m*=1, even *n*) orders and (*m*=2, odd *n*) orders, respectively. We denote the former set as the fundamental HSG (F-HSG) and the latter as the second-harmonic HSG (SH-HSG). The imaging capability is enabled by transversely scanning the sample with nanometer spatial resolution, while the generated HSG spectra, consisting of different HSG orders, are recorded by a spectrometer (Fig. 1c). Further details about the experimental setup can be found in Supplementary Materials (SM) Section S1.

The transverse spatial resolution of the proposed high-order sideband microscopy (HSM) is determined by the spot size of the NIR probe beam, since the HSG wave-mixing occurs only when the pump and probe pulses overlap in both space and time. In our experiments, we collinearly focus the MIR pump and NIR probe beams through an objective lens into silicon micro-resonators, with full-width-at-half-maximum (FWHM) transverse beam sizes being ~60 μm and ~1 μm, respectively. When imaging using the F-HSG orders, the transverse spatial resolution of our HSM is ~1.35 μm (FWHM intensity), which is determined by a knife-edge measurement across a sharp flat edge of a silicon thin film. Sub-micrometer super-resolution of ~920 nm is achieved by the SH-HSG orders, which can be attributed to the nonlinear process involving $m$=2 NIR photons (see SM Fig. S2).

**Tomographic imaging of near-field wave patterns inside a dielectric material.** The entire HSG spectrum, spanning more than one octave in frequency with more than 9 different harmonic orders (Fig. 1d), provides rich information about the near-field wave patterns under exploration. Experimental images obtained using various HSG orders are shown in the upper panel of Fig. 2a, where a silicon micro-resonator with a radius of 5.6 μm and a thickness of 1.5 μm is employed. The polarizations of the pump and probe pulses are both aligned along $x$, which allows us to measure the $x$-polarized near-field components (see below). Most intriguingly, distinct imaging patterns emerge for different HSG orders. Specifically, the images from the (1, 2) and (2, -3) orders exhibit a strong intensity node at the cylinder's center, while those from the (2, -1), (2, 1) and (2, 3) orders display weaker intensities at the center and stronger fringes along $y$. Moreover, for the (1, 4) order, the image exhibits a distinctive double-peak structure across the cylinder center along $x$. Understanding these different imaging patterns involves considering phase matching and collective accumulation and propagation of the HSG signals along $z$ within the thick dielectric resonator.

We now establish a scheme to reconstruct the 3D near-field patterns inside the resonator from the images obtained with different HSG orders. In the experiment, we intentionally control the pump and probe intensity to ensure that the HSG process remains in the perturbative region, which is characterized by the relationship: $I_{\mathrm{HSG}}^{(m,n)} \propto |E^{\mathrm{Pr}}|^{2m} \cdot |E^{\mathrm{NF}}|^{2|n|}$ (see SM Fig. S3). Here, $I_{\mathrm{HSG}}^{(m,n)}$ is the ($m$, $n$) order HSG intensity, and $E^{\mathrm{Pr}}$ and $E^{\mathrm{NF}}$ represent the probe laser field and local near-field, respectively. Consequently, the nonlinear polarization inside the silicon resonator at the coordinate ($x$, $y$, $z$) associated with the ($m$, $n$) HSG order can be described by

$$P_{\mathrm{HSG}}^{(m,n)}(x,y,z) = \varepsilon_0 \chi_{(m,n)} \cdot \; [E^{\mathrm{Pr}}(x,y,z)]^m \cdot [S^{\mathrm{NF}}(x,y,z)]^{|n|}, \tag{2}$$

where $\varepsilon_0$ is the vacuum permittivity, $\chi_{(m,n)}$ is the nonlinear susceptibility, and $S^{\mathrm{NF}}$ represents the MIR near-field ($E^{\mathrm{NF}}$) and its complex conjugate: $S^{\mathrm{NF}} = \begin{cases} E^{\mathrm{NF}}, & (n>0) \\ (E^{\mathrm{NF}})^*, & (n<0) \end{cases}$. The HSG intensity [$I_{\mathrm{HSG}}^{(m,n)}$] emitted from a specific point ($x$, $y$) thus results from the coherent integration of the nonlinear polarizations within the probe beam spot and at different $z$, which can be calculated by

$$I_{\mathrm{HSG}}^{(m,n)}(x,y) = \left| \int_0^h dz \;\, g_{(m,n)}(z) \cdot [S^{\mathrm{NF}}(x,y,z)]^{|n|} \right|^2. \tag{3}$$

Here, $h$ is the material thickness, and $g_{(m,n)}(z)$ serves as a “gate function” that accounts for the probe-field distribution and the order-dependent multiple reflections of the HSG fields. The detailed derivation of $g_{(m,n)}(z)$ is provided in SM Section S4, in which we incorporate a Gaussian transverse profile for the probe field, and consider the phase matching process, as well as the transmission properties of the HSG signals generated at $z$ within the silicon resonator.

We observe that the gate function of different HSG orders exhibit distinct $z$-dependence due to the phase sensitivity inherent to the process, making them suitable for resolving the near-fields at different depths. In Fig. 2b, we plot the real part of $g_{(m,n)}(z)$ and its integration $\int_0^z g_{(m,n)}(z')\, dz'$.

For (1, 2), HSG is more sensitive to the near-fields at $z \approx 0.75$ μm, diminishing to almost zero sensitivity at $z \approx 0.5$ and 1.25 μm. This sensitivity pattern is compensated by (1, 4), which exhibits higher sensitivity at $z \approx 0.5$ and 1.25 μm. For (2, -3), the gate function is more sensitive to the field close to the interface with the substrate at $z=0$. The integration of the gate function over the full height of the disk shows a value close to zero for (1, 4), while a finite brightness for the other two orders. These results qualitatively explain the intensity variations for different orders at the cylinder's center (Fig. 2a).

The distinct sensitivity of various HSG orders to the local near-fields at different depths, facilitated by the gate function, provides a unique opportunity to realize 3D tomographic reconstruction of near-field wave patterns deep inside the silicon micro-resonator. To achieve this, we initially model the HSG images of different orders ($m$, $n$) using Eq. (3) to obtain the modeled HSG images $I_{\mathrm{model}}^{(m,n)}(x, y)$, incorporating an initial guess for the near-field distribution $E^{\mathrm{NF}}(x, y, z)$. Here, the near-field distribution along $z$ is parameterized by second-order polynomials: $E^{\mathrm{NF}}(x, y, z) = \alpha_0(x, y) + \alpha_1(x, y) \cdot z + \alpha_2(x, y) \cdot z^2$. A global optimization algorithm is then implemented to minimize the difference between the modeled images $[I_{\mathrm{model}}^{(m,n)}(x, y)]$ and the experimental images $[I_{\mathrm{expr}}^{(m,n)}(x, y)]$ of different harmonic orders, to determine the complex-valued coefficients $\alpha_0(x, y)$, $\alpha_1(x, y)$ and $\alpha_2(x, y)$. Further details about the reconstruction algorithm can be found in SM Section S5.

We present the reconstructed HSG images for different ($m$, $n$) in the lower panel of Fig. 2a, which exhibit excellent agreement with the experimental results. For a comprehensive comparison, the fitting results across the center of the silicon cylinder along $y$ are plotted in Fig. 2c. In Fig. 2d, we compare the experimentally reconstructed distribution of $|E^{\mathrm{NF}}|^2$ at various depths with that obtained by the finite-element-method (FEM) simulations (see SM Section S3). Quantitative agreement between the two results is noted. Furthermore, the lower panel of Fig. 2d presents a *y*-*z* cross-sectional view for a comprehensive field distribution comparison across the cylinder center. In SM Movie S1, we provide a video presenting a slice-view comparison of the near-field distribution obtained from both the experimental reconstruction and the FEM simulations. Overall, the experimentally reconstructed patterns are in good agreement with the FEM simulations.

The phase sensitivity inherent in our approach along the longitudinal direction enables subwavelength super-resolution imaging along $z$ [55,56]. In this study, by employing 6 different HSG orders to determine 5 coefficients (including both the real and imaginary parts of $\alpha_1$ and $\alpha_2$), the fitting problem becomes overdetermined. This allows us to retrieve of a unique set of coefficients with high precision, when the experimental noise is ignored. The longitudinal spatial resolution is, on the other hand, limited by the experimental signal-to-noise ratio (SNR). Our numerical analysis shows that, with the experimental SNR of different HSG orders considered, we can achieve a high longitudinal spatial resolution of ~130 nm (see SM Section S7). However, it is important to note that the high longitudinal spatial resolution in this study is conditional. With a more complex longitudinal near-field distribution, more coefficients under the polynomial or other bases would be necessary, which would require more HSG orders considered in the algorithm to maintain the overdetermined condition.

**Polarization-resolved near-field imaging.** As the excited near-field wave patterns can possess different polarization components, understanding which polarization component is mapped by HSM is crucial. For a normal incident probe beam along $z$ (Fig. 1c), we can assume that $E_z^{\mathrm{Pr}} \ll E_{x,y}^{\mathrm{Pr}}$, where $E_{x/y/z}^{\mathrm{Pr}}$ represents the probe-field components polarized along $x$, $y$ and $z$ directions. The HSG polarization in Eq. (2) is expanded into a matrix form for different polarizations. Taking the (2, -1) order SH-HSG as an example, the HSG polarization is given by (see SM Section S8)

$$\begin{pmatrix} P_x^{(2,-1)} \\ P_y^{(2,-1)} \\ P_z^{(2,-1)} \end{pmatrix} = \varepsilon_0 \chi^{(3)} \begin{pmatrix} 3(E_x^{\mathrm{Pr}})^2 (E_x^{\mathrm{NF}})^* + 2E_x^{\mathrm{Pr}} E_y^{\mathrm{Pr}} \left(E_y^{\mathrm{NF}}\right)^* + \left(E_y^{\mathrm{Pr}}\right)^2 (E_x^{\mathrm{NF}})^* \\ 3\left(E_y^{\mathrm{Pr}}\right)^2 \left(E_y^{\mathrm{NF}}\right)^* + 2E_y^{\mathrm{Pr}} E_x^{\mathrm{Pr}} (E_x^{\mathrm{NF}})^* + (E_x^{\mathrm{Pr}})^2 \left(E_y^{\mathrm{NF}}\right)^* \\ \left(E_y^{\mathrm{Pr}}\right)^2 (E_z^{\mathrm{NF}})^* + (E_x^{\mathrm{Pr}})^2 (E_z^{\mathrm{NF}})^* \end{pmatrix}. \tag{4}$$

Consequently, the *x*-component of the MIR near-field ($E_x^{\mathrm{NF}}$) can be resolved by detecting the intensity of the *x*-polarized HSG radiation ($\left|P_x^{(2,-1)}\right|^2$), while placing the probe-beam polarization also along *x* ($E_x^{\mathrm{Pr}}$). Following the same principle, the *y*-component ($E_y^{\mathrm{NF}}$) can be resolved by jointly rotating the probe-beam polarization and the HSG polarizer (see SM Fig. S1) to align with the *y* direction.

We characterize the polarization-resolving capability of HSM by scanning the relative angle $\theta$ between the probe-beam polarization and the MIR-pump polarization, while maintaining the HSG polarizer for detection along the same direction as the probe-beam polarization (inset of Fig. 3a). To avoid complex polarization distribution in this characterization, we intentionally excite a 1.5-μm thick silicon film. The experimental results for the (2, -1) order SH-HSG are shown in Fig. 3a, which conforms to the $(\cos\theta)^2$ function, in excellent agreement with Eq. (4) (see SM Section S8). The polarization extinction ratio can reach ~30 dB.

Polarization distribution of the near-fields in a finite-sized resonator becomes complex, serving as an ideal platform to test the polarization-resolving capability of HSM. Figure 3b depicts the FEM simulated vector near-field distribution in a silicon resonator with a radius of 5.1 μm and a thickness of 0.5 μm. Although the polarization of the incident MIR pump aligns along *x*, the excitation of the electric and toroidal dipole moments leads to both the $E_x^{\mathrm{NF}}$ and $E_y^{\mathrm{NF}}$ components with distinct spatial distributions. The experimental images of the $E_x^{\mathrm{NF}}$ and $E_y^{\mathrm{NF}}$ components, obtained using the (2, -1) order, are shown in Fig. 3c and d, respectively. The results can be well reproduced by the simulations using Eq. (3) and considering the near-field distribution calculated by the FEM simulations, as shown in Fig. 3e and f.

It is noteworthy that other HSG orders also exhibit polarization sensitivity for near-field microscopy. Specifically, we observe that the SH-HSG orders generally exhibit a high polarization extinction ratio of ~30 dB, whereas the F-HSG orders have extinction ratios about two orders of magnitude lower (~10 dB). This discrepancy can be attributed to the different symmetries inherent in these two types of nonlinear optical conversion (see SM Section S8).

**Phase-resolved near-field imaging through quantum-path interference.** Finally, we demonstrate the capability of HSM to resolve the phase of local near-fields. This is achieved through quantum-path interference (QPI) by overlapping neighboring HSG orders in the spectrum[53]. In the experiment, this overlap is accomplished by broadening the probe-pulse spectrum through supercontinuum generation (see SM Section S1).

As illustrated in Fig. 4a, when the probe-pulse bandwidth exceeds $\frac{2\omega_{\mathrm{Pu}}}{m}$, two quantum paths associated with neighboring HSG orders [e.g. ($m$, $n$) and ($m$, $n$-2)] emerge, connecting the same initial and final states, with the final-state frequency given by $\omega = m\omega_1 + n\omega_{\mathrm{Pu}} = m\omega_2 + (n-2)\omega_{\mathrm{Pu}}$. Here, $\omega_1$ and $\omega_2$ respectively denote two probe-light frequencies within the bandwidth, with $\omega_2 - \omega_1 = \frac{2\omega_{\mathrm{Pu}}}{m}$, and their corresponding phases are $\phi_1$ and $\phi_2$. In this demonstration, we take the SH-HSG orders (2, 1) and (2, -1) as an example, and $\omega_1$ and $\omega_2$ are, thus, separated by a pump-laser frequency $\omega_{\mathrm{Pu}}$ (Fig. 4a).

When the probe pulse is chirped by second-order dispersion and the time delay between the pump and probe pulses ($\tau$) is fixed, QPI leads to stable spectral modulation that carries essential information about the near-field phase, $\phi_{\mathrm{NF}}$. Specifically, the frequency for the constructive spectral interference at position ($x$, $y$) is given by $\omega(x,y) = 2\omega_c + 2\frac{\phi_{NF}(x,y) - l\cdot\pi + \phi_\tau}{D_2\cdot\omega_{Pu}}$, where $l$ is an integer, $\omega_c$ and $D_2$ are the center frequency and the group-delay dispersion (GDD) of the probe pulse, respectively, $\phi_{NF}(x, y)$ represents the local near-field phase at ($x$, $y$) and $\phi_\tau = -\omega_{Pu}\tau$ is a delay-related static phase. (Detailed derivation is provided in SM Section S9.) Consequently, the HSG spectrum in the overlapping regions exhibits intensity modulations with the bright stripes separated by $\Delta\omega = \frac{2\pi}{D_2\cdot\omega_{Pu}}$. Whenever $\phi_{NF}$ varies by ±π, the bright stripes in the spectrum shift by $\pm\Delta\omega$, corresponding to a shift of $\Delta l = \pm 1$.

In Figs. 4c and d, we present the experimentally measured spatial variations of the spectral interference pattern in the overlapping regions of the (2, 1) and (2, -1) orders by scanning the probe beam transversely across the silicon resonator along two trajectories (i) and (ii) (illustrated in Fig. 4b), respectively. The measurements are conducted for the $y$-polarized near-field components ($E_y^{NF}$). The radius and thickness of the silicon cylinder are 5.1 μm and 0.5 μm, respectively, in this case. A thin disk is chosen for this demonstration to avoid the thickness-related averaging effect on $\phi_{NF}$. The results clearly show a collective shift of the bright spectral stripes by $\Delta l = \pm 1$ in Fig. 4c and d, indicating a π phase shift of $\phi_{NF}$ along the trajectories.

The spatial variations of $\phi_{NF}$ can be quantitatively extracted, as shown in Fig. 4g, by considering the probe-pulse GDD, $D_2$, and the pump-pulse frequency, $\omega_{Pu}$, both of which are determined experimentally (see SM Section S9). In the inset of Fig. 4g, we present the distribution of the $y$-polarized near-field phase obtained from the FEM simulations, which exhibits π phase shift across each quarter of the pillar. By considering QPI between the HSG orders, we can numerically simulate the shifts of the interference patterns along the same trajectories (i) and (ii), as shown in Figs. 4e and f, which exhibit excellent agreement with the experimental results (Figs. 4c and d). The lineouts extracted from the simulation results are compared directly with the experimental results in Fig. 4g for the *y*-polarized components. Notably, for the *x*-polarized near-fields, our experiments show the absence of such a such a π-phase shift (see SM Fig. S12D), which is also in excellent agreement with the FEM simulation results. These results unequivocally demonstrate the capability of our method to comprehensively characterize vector near-fields inside the silicon resonator, including spatial distribution of both phase and polarization states.

Here, we note that it is promising to reconstruct depth-dependent near-field phases in a thicker resonator by experimentally analyzing the interference patterns in the overlapping regions between multiple neighboring HSG order pairs. While the current experimental conditions limit the clarity of interference patterns due to the low intensities of higher-order harmonics, advancements in experimental setups and detection sensitivity could significantly improve phase retrieval. Additionally, a more sophisticated reconstruction algorithm that accounts for quantum-path interference is required for accurate phase retrieval.

## Discussion and Conclusion

In this work, we demonstrate nanometer-resolution, 3D tomographic and vectorial near-field imaging deep inside dielectric materials, achieving comprehensive characterization of the anapole modes excited within silicon-cylindrical optical resonators. We first note that special care is required when the HSG photons are on resonance with interband excitations, as this introduces wavelength-dependent variations in the material's linear refractive index. Under such resonant conditions, it is necessary to determine the refractive indices either through experimental measurements or by consulting reference data (see SM Section S6).

Furthermore, our approach utilizes the HSG method to assess the depth distribution of MIR near-fields through the broad HSG spectrum generated by varying the harmonic orders *m* and *n*. A similar concept can be adapted to probe near-fields at shorter wavelengths, such as visible or ultraviolet. However, in these cases, the use of shorter-wavelength probe fields shifts the high-order harmonic photon energy into the extreme ultraviolet range, posing significant challenges for detection. As indicated by Eq. (1), the same longitudinal spatial resolution can be achieved by varying the probe-laser wavelength ($\lambda_{\mathrm{pr}}$), with the required spectral bandwidth $\delta\lambda_{\mathrm{pr}}$ scaling proportionally to $\lambda_{\mathrm{pr}}^2$. The stronger refractive-index dispersion at shorter wavelengths further results in larger $B_{(m,n)}$ values, thereby reducing the spectral bandwidth required for tomographic applications (see the discussions in Methods).

In practice, a wavelength-tunable probe laser offers an alternative means to extract near-field information at various depths within dielectric nano-resonators. We demonstrate this concept through a combined experimental and theoretical investigation of the near-field distribution within a 430 nm-thick free-standing $Si_3N_4$ nano-resonator, resonantly excited by a 1.6 μm pump laser and probed with a tunable laser varying from 263 nm to 343 nm. Distinct patterns revealing near-field distributions at different depths are observed (see SM Section S10). For phase-resolved measurements, excitation and detection of two adjacent harmonic orders are required.

Finally, the high-order nonlinear optical microscopy technique demonstrated here holds great promise for applications across a variety of material platforms and other laser-induced MIR wave phenomena beyond resonant near-fields. These include hyperbolic phonon-polaritons[57,58], topological photonic edge states[59,60], among others. While previous investigations have focused on characterizing the surface or depth-averaged near-field properties of these unique wave phenomena, our method offers the potential for comprehensive 3D imaging of their distribution and propagation, which will substantially advance our understanding and manipulation of these phenomena.

**Acknowledgments:** This work was accomplished in Fudan University. We wish to thank Chuanshan Tian for technical support. Z. T. also acknowledges the support from the National Key Research and Development Program of China (Grant No. 2021YFA1400200) and the National Natural Science Foundation of China (No. 12274091). L. Z. and Z. T. acknowledge the support from the National Key Research and Development Program of China (Grant No. 2022YFA1404700). L. Z. and Z. T. acknowledge the support from the National Natural Science Foundation of China (Grant No. 12221004). L. Z. and Z. T. acknowledge the support from the National Natural Science Foundation of China (Grant No. 62192771). Z. T. acknowledge the support from the Shanghai Municipal Science and Technology (Grant No. 22JC1400200 and No. 23dz2260100).

**Author contributions:**

Conceptualization: Z. T., L. Z.

Experiment: B. Z., Q. C., Y. L. S. Z.

Theory: B. Z., Q. C., Q. H.

Experimental support: Y. L., S. Z., W. L.

Supervision: Z. T., L. Z.

Writing – original draft: B. Z., Q. C., L. Z., Z. T.

Writing – review & editing: Y. L., S. Z., W. L., Q. H.

**Competing interests:** Authors declare that they have no competing interests.

**Inclusion & ethics statement:** All collaborators of this study have fulfilled the criteria for authorship required by Nature Portfolio journals have been included as authors, as their participation was essential for the design and implementation of the study. Roles and responsibilities were agreed among collaborators ahead of the research. This work includes findings that are locally relevant, which have been determined in collaboration with local partners. This research was not severely restricted or prohibited in the setting of the researchers, and does not result in stigmatization, incrimination, discrimination or personal risk to participants. Local and regional research relevant to our study was taken into account in citations.

**Data and materials availability:** All data are available in the main text or the supplementary materials.

**Code availability:** The codes used for our analyses are available upon request to the corresponding author, Z. T.

**Supplementary Materials**

Supplementary Text

Figs. S1 to S18

Tables S1 to S2

References (*61–71*)

Movie S1

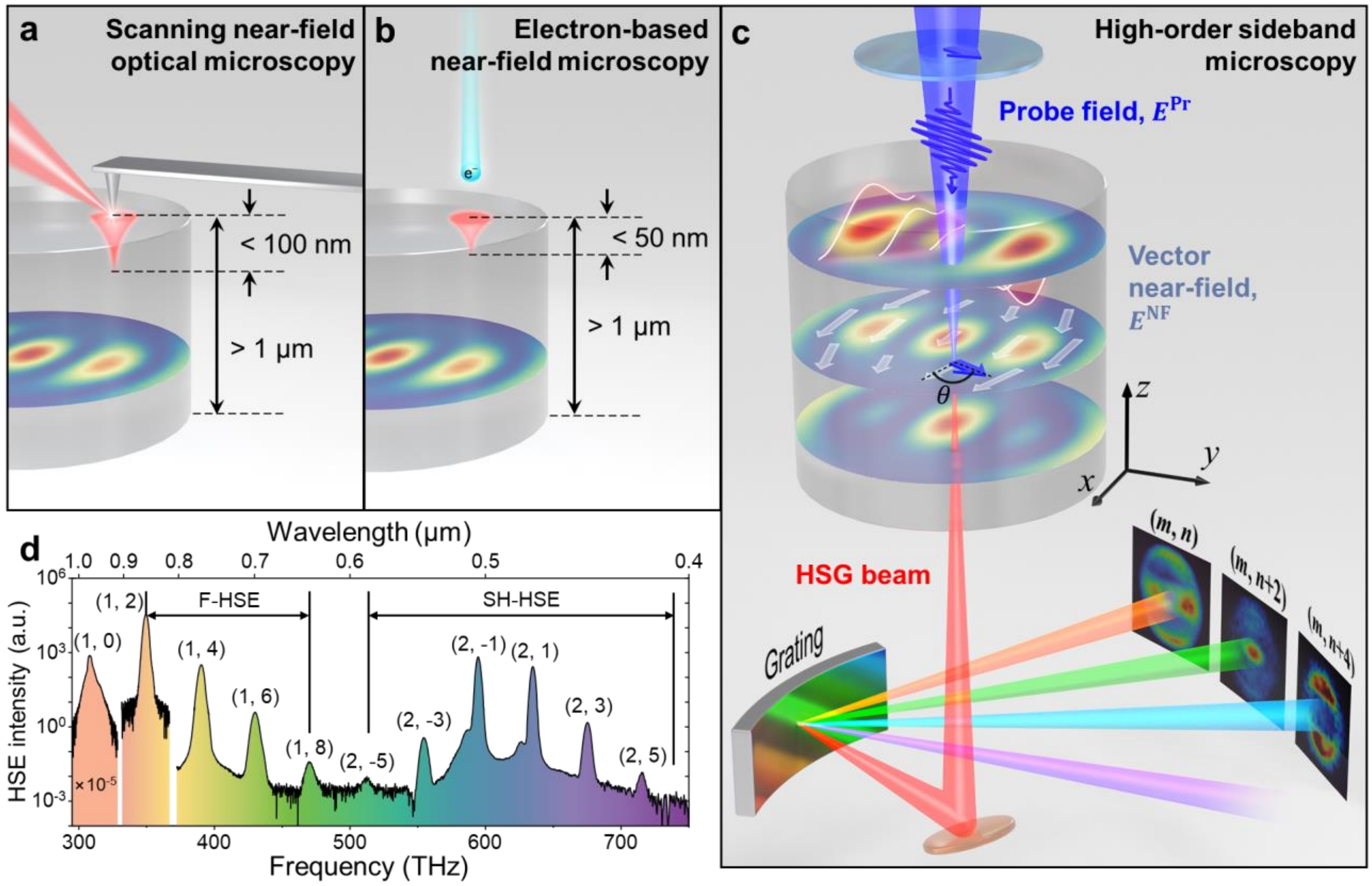

**Figure 1. Comparison of near-field imaging techniques. a.** Illustration of scanning near-field optical microscopy probing the near-field distribution inside a micrometer-thick dielectric optical resonator. The probing depth is typically <100 nm. **b.** Illustration of electron-based near-field microscopy. The probing depth is typically <50 nm. **c.** Schematics of the experimental setup for high-order sideband microscopy. The probe-field $E^{\mathrm{Pr}}$ is focused within the dielectric optical resonator, inducing HSG through the nonlinear wave-mixing with the local near-field ($E^{\mathrm{NF}}$). The relative polarization angle between $E^{\mathrm{Pr}}$ and $E^{\mathrm{NF}}$ is denoted as $\theta$. HSG beam is generated through high-order nonlinear wave mixing between $E^{\mathrm{Pr}}$ and $E^{\mathrm{NF}}$, which consists of different HSG orders: $\omega_{(m,n)} = m\omega_{\mathrm{Pr}} + n\omega_{\mathrm{Pu}}$. Near-field images of different HSG orders are simultaneously recorded through a spectrometer. **d.** Typical experimental HSG spectrum from a silicon optical resonator. The MIR pump wavelength is $\lambda_{\mathrm{Pu}}$≈13.4 μm.

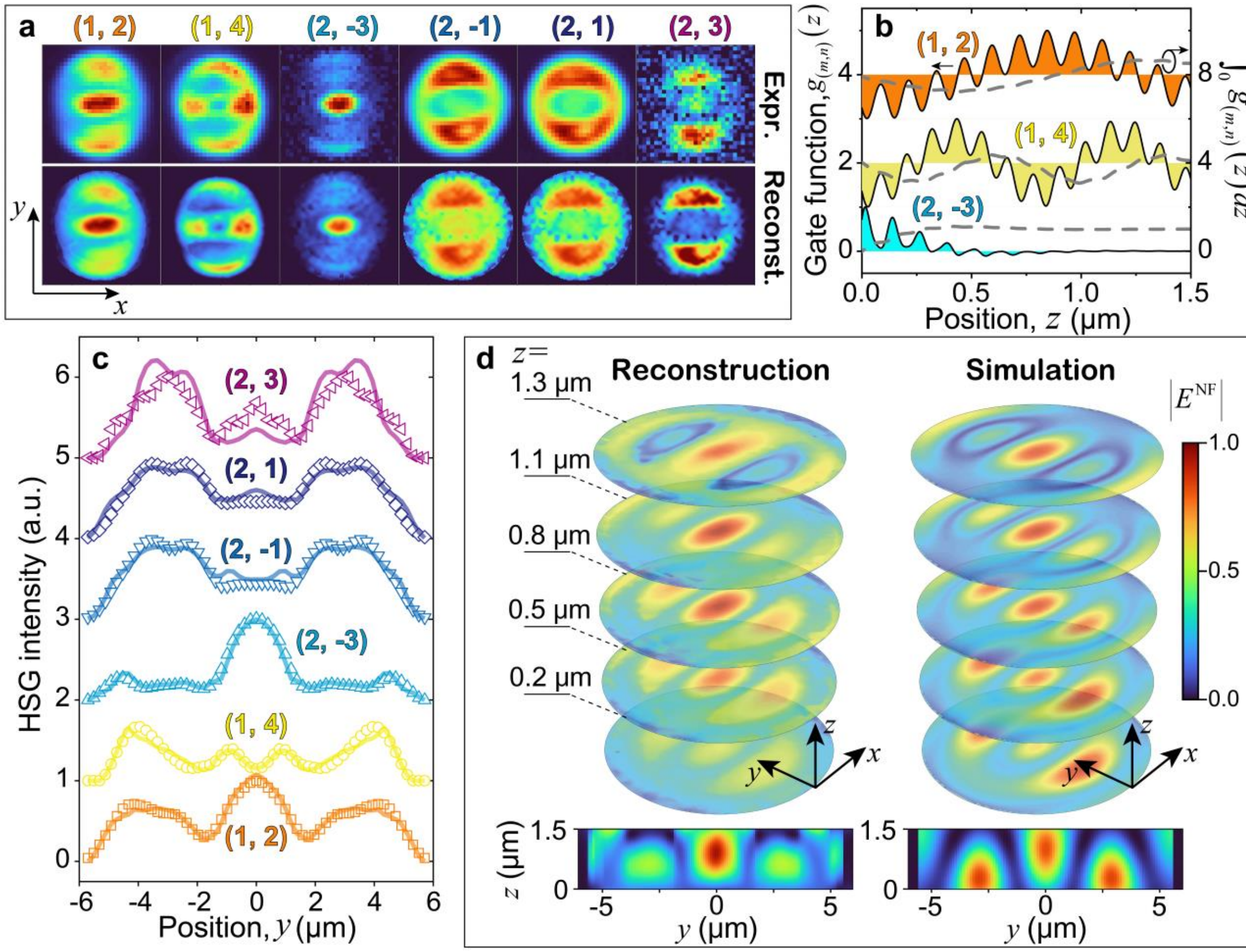

**Figure 2. Tomographic imaging of near-field wave patterns. a.** Experimental (upper panel) and reconstructed (lower panel) near-field images using different HSG orders. **b.** Real part of gate function $g_{(m,\,n)}$ and its integration [$\int_0^z g_{(m,n)}(z)dz$)] as a function of the depth *z* for different HSG orders. **c.** Fitting results of the cross-sectional HSG images at *x*=0 along *y*. **d.** (Left) *x*-*y* cross-sectional images of near-field strength ($|E^{\mathrm{NF}}|$) at different *z* obtained by the tomographic reconstruction of HSG images shown in **a.** (Right) *x*-*y* cross-sectional images at different *z* obtained using the FEM simulation. *y*-*z* cross-sectional views of near-field distributions across the cylinder center are shown in the lower panels for both experimental reconstruction and FEM simulation.

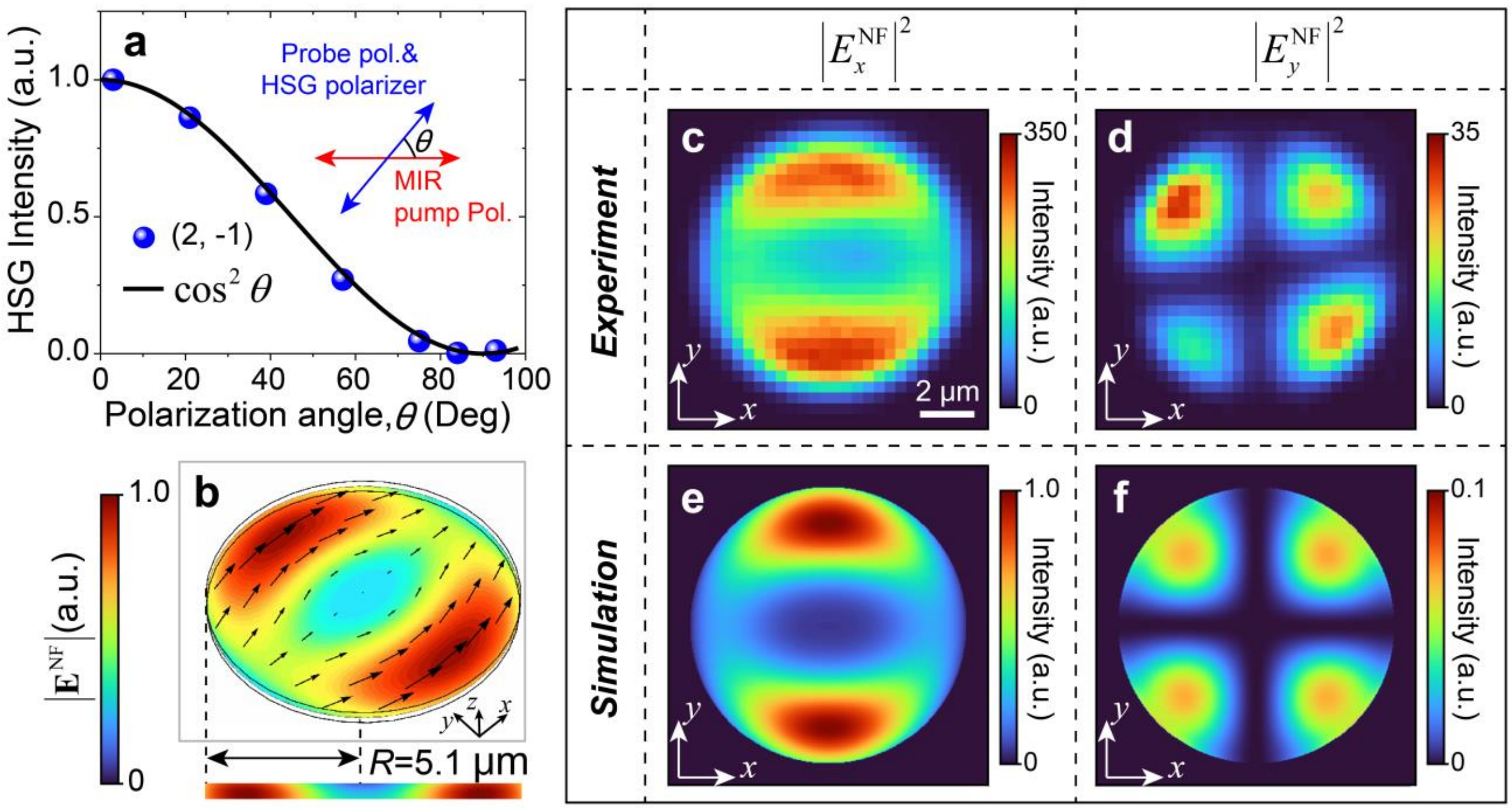

**Figure 3. Polarization-resolved near-field imaging. a.** Intensity of the HSG order (2, -1) as a function of polarization angle $\theta$, measured on a 1.5-μm thick silicon film excited by linearly polarized pump beam. The solid line represents the fitting result using the function $\cos^2\theta$. **Inset:** Illustration of the definition of the polarization angle $\theta$. **b.** Near-field amplitude and polarization distribution inside the silicon cylinder obtained from the FEM simulation. **c.** and **d.** Experimental near-field HSG images for the *x* and *y* components, respectively, using the HSG order (2, -1). **e.** and **f.** Simulated HSG images for the *x* and *y* components, respectively, using the near-field distribution obtained from the FEM simulations.

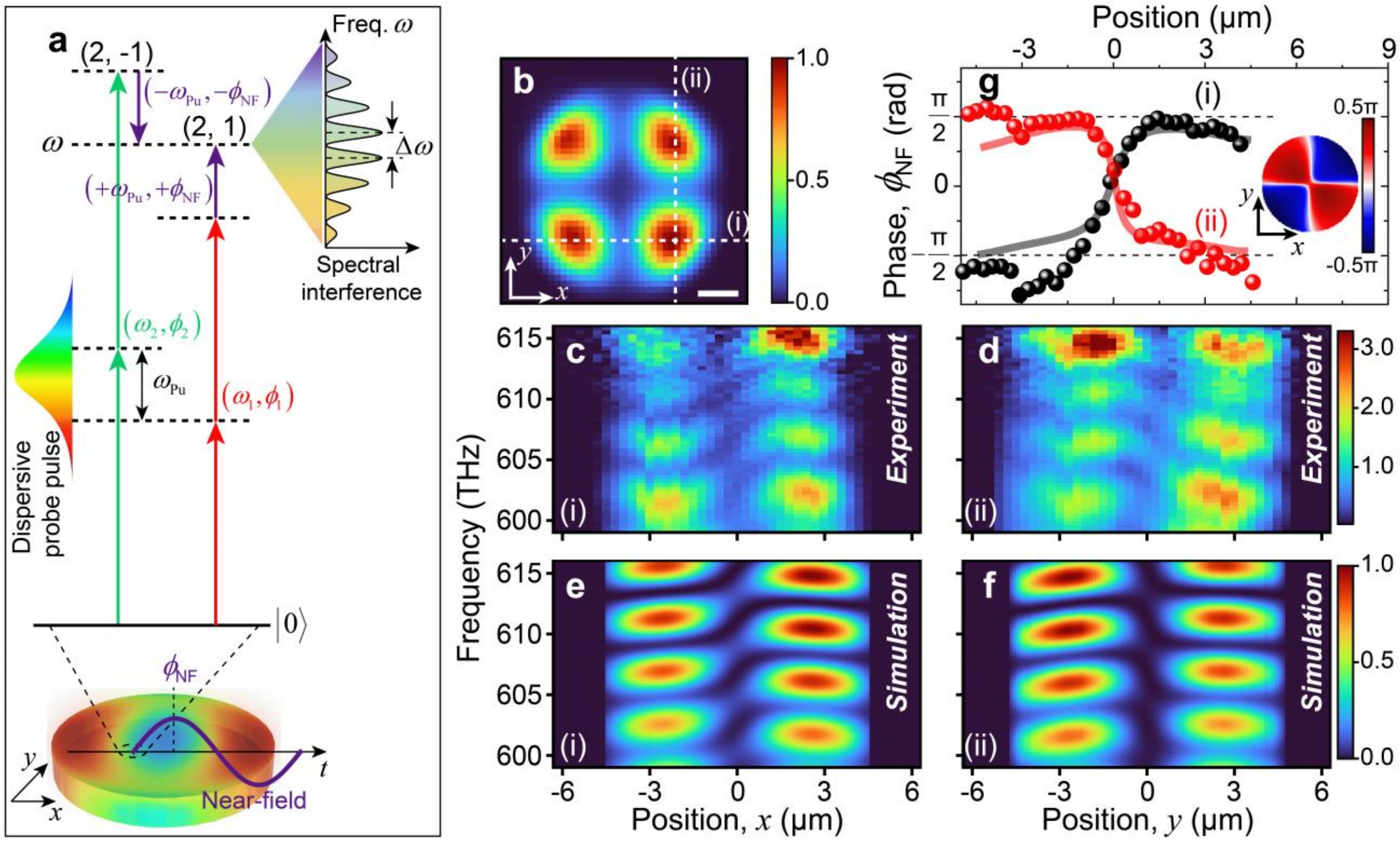

**Figure 4. Phase-resolved near-field imaging. a.** Illustration of quantum-phase interferometry, which provides phase-resolving capability. **b.** Near-field HSG imaging of the *y*-component field using HSG order (2, -1). The scale bar indicates 2 μm. White dashed lines label the trajectories (i) and (ii) for **c** - **f** below. **c.** and **d.** Experimentally measured spatial variation of the spectral interferogram along the trajectories (i) and (ii), respectively. **e.** and **f.** Simulated results along (i) and (ii) in direct comparison with **c** and **d**. **g.** Near-field phase, $\phi_{NF}$, as a function of position along (i) and (ii) extracted from **c - f**. Symbols represent experimental results from **c** and **d**, while solid lines are the simulation results from **e** and **f**. **Inset:** Spatial distribution of $\phi_{NF}$ obtained from the FEM simulation.

**Methods Summary**

**Experimental setup and sample fabrication.** The experimental setup is illustrated in SM Fig. S1. The MIR pump and NIR probe beams were collinearly focused by an objective lens (Obj. lens) with a numerical aperture (NA) of 0.28. This configuration resulted in a FWHM beam size of ~1 μm for the probe laser, while the FWHM beam size of the pump laser was ~60 μm. The polarization direction of the probe pulse was adjusted by a NIR half-waveplate. The intensity of the pump pulse was adjusted by a pair of MIR polarizers. The generated HSG beam was collected by another objective lens (Obj. lens, NA=0.40), and was recorded by a spectrometer (Horiba iHR320), after passing through a broadband polarizer. The samples were mounted on a 3D piezo-stage. The real-space imaging of near-field wave patterns was achieved by transversely scanning the sample in the x-y plane with a precision of ~5 nm. In most experiments, the step sizes in the transverse (x and y) directions ranged from 0.3 to 0.4 μm. For the measurement of near-field phases, the spectrum of the NIR probe pulse was slightly broadened by focusing the probe beam into a 4 mm-thick YAG crystal, leading to supercontinuum generation. The silicon-cylindrical optical resonators were prepared by using Plasma Enhanced Chemical Vapor Deposition and Inductively Coupled Plasma Etching. The detailed fabrication process is provided in SM Section S2.

**Mechanisms for tomographic imaging.** The wavevector mismatch associated with the $(m, n)$-th harmonic order is given by

$$\Delta k_{(m,n)} = k_{(m,n)}\left(n\omega_{\mathrm{pu}} + m\omega_{\mathrm{pr}}\right) - m \cdot k_{\mathrm{pr}}\left(\omega_{\mathrm{pr}}\right) - n \cdot k_{\mathrm{pu}}\left(\omega_{\mathrm{pu}}\right). \quad (5)$$

In our analysis, the spectral bandwidth in the experiments is varied either by varying the harmonic orders ($n \rightarrow n + \delta n$) or by changing the probe laser frequency ($\omega_{\mathrm{pr}} \rightarrow \omega_{\mathrm{pr}} + \delta\omega_{\mathrm{pr}}$). For simplicity, we assume that the number of probe photons ($m$) remains constant. The variation of the wavevector mismatch can be approximated as:

$$\delta k_z = \frac{1}{v_g^{(m,n)}} d\omega_{(m,n)} - \frac{m}{v_g^{\mathrm{pr}}} \delta\omega_{\mathrm{pr}} - \delta n \cdot k_{\mathrm{pu}}, \tag{6}$$

where $v_g^{(m,n)}$ and $v_g^{\mathrm{pr}}$ represents the group velocities at the wavelengths $\lambda_{(m,n)}$ and $\lambda_{\mathrm{pr}}$, respectively, and $d\omega_{(m,n)} = \delta n \cdot \omega_{\mathrm{pu}} + m \cdot \delta\omega_{\mathrm{pr}}$. We further define $\delta\omega_{(m,n)} = \delta n \cdot \omega_{\mathrm{pu}}$, representing the output frequency variation solely induced by changing the harmonic orders *n*. Hence, we can write the variation in the wavevector mismatch as:

$$\delta k_z = \frac{2\pi\left[n_g^{(m,n)} - n_{\mathrm{pu}}\right]}{\lambda_{(m,n)}^2} \delta\lambda_{(m,n)} + m \frac{2\pi\left[n_g^{(m,n)} - n_g^{\mathrm{pr}}\right]}{\lambda_{\mathrm{pr}}^2} \delta\lambda_{\mathrm{pr}}, \tag{7}$$

where $n_g^{(m,n)}$ and $n_g^{\mathrm{pr}}$ represent the group refractive indices of the dielectric medium at $\lambda_{(m,n)}$ and $\lambda_{\mathrm{pr}}$, respectively, and $n_{\mathrm{pu}}$ is the refractive index at $\lambda_{\mathrm{pu}}$. Furthermore, by defining the coefficients:

$$\begin{cases} A_{(m,n)} = 2\pi\left[n_g^{(m,n)} - n_{\mathrm{pu}}\right] \\ B_{(m,n)} = 2\pi\left[n_g^{(m,n)} - n_g^{\mathrm{pr}}\right] \end{cases}, \tag{8}$$

Eq. (7) is alias to Eq. (1) in the main text.

When the pump laser wavelength is in the mid-infrared range, the probe laser wavelength is near-infrared, and the imaging takes only the lowest-order nonlinear optical process (e.g. $m$=1, $n$=2), we have $\lambda_{(m,n)} \approx \lambda_{\mathrm{pr}}$. In this case, the $B(m,n)$ coefficient becomes very small, which would require an extremely broad $\lambda_{\mathrm{pr}}$ spectrum for high-resolution tomographic imaging. On the other hand, the large wavelength difference between $\lambda_{(m,n)}$ and $\lambda_{\mathrm{pu}}$ supports a large refractive-index difference in $A(m,n)$. This makes tomographic imaging by varying the harmonic orders ($n$) a feasible solution.

In contrast, when both the pump and probe wavelengths are short (e.g. for imaging nano-resonators at the visible and ultraviolet wavelengths), the dispersion of the refractive index at these wavelengths can yield a large $B_{(m,n)}$ coefficient. Under such conditions, tomographic imaging by varying $\lambda_{\mathrm{pr}}$ becomes a practical and effective approach.